\documentclass[fleqn,usenatbib,twocolumn]{rasti}

\usepackage{newtxtext}
\usepackage[T1]{fontenc}

\DeclareRobustCommand{\VAN}[3]{#2}
\let\VANthebibliography\thebibliography
\def\thebibliography{\DeclareRobustCommand{\VAN}[3]{##3}\VANthebibliography}

\usepackage{graphicx}	
\usepackage{amsmath}	

\usepackage{xcolor}
\usepackage{array}

\usepackage{natbib}
\usepackage{amssymb}

\defcitealias{Kostogryz2026}{K26}

\title[Forward model]{Model Choice or Model Bias? How stellar spectral libraries and activity modelling choices influence Ariel spectra — I. Forward model}

\author[Nèmec et al.]{
N.-E. Nèmec$^{1,2}$,\thanks{E-mail: nnemec@ice.csic.es}
A. Thompson$^{3}$, 
Ò. Porqueras-León$^{1,2}$,
S. Stucki$^{1,2}$  ,
N. R. C. Gomes$^{1,2}$ ,
D. Baroch $^{2}$,  \newauthor
and I. Ribas$^{1,2}$
\\
$^{1}$Institut de Ciències de l’Espai (ICE, CSIC), Campus UAB, c/ Can Magrans s/n, 08193 Bellaterra, Barcelona, Spain\\
$^{2}$Institut d’Estudis Espacials de Catalunya (IEEC), Edifici RDIT, Campus UPC, 08860 Castelldefels (Barcelona), Spain\\
$^{3}$INAF-Osservatorio Astronomico di Palermo, Piazza del Parlamento, 1, 90134 Palermo, Italy}

\begin{document}
\label{firstpage}
\maketitle

\begin{abstract}
The Ariel mission will conduct a population-level survey of approximately 1000 exoplanet atmospheres across its nominal lifetime, requiring unprecedented consistency in the treatment of stellar contamination of transmission spectra. We present a systematic comparison of stellar spectral libraries and modelling approaches for computing facular and spot contamination spectra across the full Ariel wavelength range (0.5--7.8 $\mu$m) for six stellar spectral types spanning F3 to M2. Using a surface-resolved forward model with approximately 200 randomly generated active region configurations per spectral type, we compute contamination spectra at Ariel Tier 3 resolution and compare three 1D spectral libraries against 3D MHD stellar atmosphere spectra as a physical ground truth. We find that library-dependent differences in the facular treatment introduce systematic biases exceeding the Ariel Tier 3 sensitivity threshold of approximately 20 ppm across most of the wavelength range, for most spectral types and activity levels considered. These biases align with the absorption bands of water, carbon dioxide, carbon monoxide, and methane, rendering them indistinguishable from genuine atmospheric signals at Ariel's precision. Our results demonstrate that the choice of stellar spectral library, in particular the faculae treatment is a critical and currently underappreciated source of systematic error for Ariel atmospheric characterisation. The impact on atmospheric retrievals is quantified in a companion paper.
\end{abstract}

\begin{keywords}
keyword1 -- keyword2 -- keyword3
\end{keywords}

\section{Introduction}

Since the discovery of the first exoplanet around a Sun-like star in the mid-1990s, the field has advanced from detection to detailed atmospheric characterization. Out of over 6,000 known exoplanets, the James Webb Space Telescope (JWST) has already probed the atmospheres of roughly 100 targets using transmission spectroscopy \citep{Seager_2000, Charbonneau2002, Tinetti2007, Yan2019, Rustamakulov2023}. As a planet transits its host star, wavelength-dependent variations in atmospheric opacity alter the observed transit depth, encoding information about atmospheric composition. However, disentangling planetary signatures from stellar spectra remains challenging due to stellar surface heterogenieties, brought about by stellar magnetism. 
Occulted features crossed by the planet leave distinct imprints in transit light curves \citep[e.g.,][]{PetitditdelaRoche2024, Ahrer2025, Fournier-Tondreau2025, Radica2025}, whereas unocculted active regions drive the Transit Light Source Effect \citep[TLSE;][]{Rackham2018}. Standard corrections rely on disc-averaged representations of active regions \citep{Rackham2018, Rackham2019} or joint atmospheric retrievals \citep{Pinhas2018, Iyer2020, Thompson2024}. Additional stellar biases, such as fixing limb-darkening coefficients or adopting inadequate limb-darkening laws, can introduce errors of $\sim$3\% in inferred planetary radii \citep{Espinoza2015, Espinoza2016} and wavelength-dependent biases exceeding 100 ppm \citep{Coulombe2024, Keers2024}.
The Ariel mission \citep{Tinetti2018,Tinetti2022}, launching in the early 2030s, will conduct a population-level survey of $\sim$1000 transiting exoplanets. Ariel's tiered observing strategy involves repeated visits to high-priority targets, making epoch-to-epoch consistency essential. Drawing lessons from JWST is vital to fully leverage these multi-visit observations. Recently, \cite{Thompson2026} demonstrated that uncorrected inter-visit stellar variability introduces significant biases in retrieved atmospheric parameters. To balance computational efficiency, scalability, and interpretability, they adopt a simplified 1D stellar activity model that is widely used in the retrieval context. Spots and faculae are treated as separate, distinct components, with each active region represented by a spectral energy distribution (SED) from the \texttt{PHOENIX BT-Settl} library \citep{Allard2014} characterised by its temperature contrast. However, such models neglect center-to-limb variations (CLV) and cannot reproduce the more complex temperature and spectral structure of faculae. As shown by \cite{Nemec2026} in the solar context, disc-averaged models fail to reproduce true contamination signals, underscoring the need for physically motivated facular models incorporating CLV.
Recent 3D radiative magnetohydrodynamic (MHD) simulations combining \texttt{MURaM} \citep{Voegler2005, Rempel2014} and \texttt{MPS-ATLAS} \citep{Witkze2021} offer first-principles models of stellar atmospheres that account for interactions among convection, magnetic fields, and radiation \citep[][hereafter K26]{Kostogryz2026}. Applying magnetic fields above 100 G (associated with faculae) reveals that faculae brighten hot F-type stars but weaken toward cooler stars, even appearing dark in the coolest models \citepalias{Kostogryz2026}. Furthermore, standard 1D and 3D CLV coefficients in forward models can produce transit depth discrepancies up to 3\% in common photometric bands (Nèmec et al., submitted).
In this work, we investigate how different treatments of faculae and choices of stellar atmosphere models propagate into transmission spectra. We compare simple TLSE prescriptions against surface-resolved active region modeling using \texttt{StarSim} \citep{Herrero2016, Rosich2020} across publicly available spectral libraries, focusing on Ariel's full spectral range (0.5–7.8 $\mu$m) at Tier 3 resolution. Section 2 describes the spectral libraries, Section 3 details the spatial modeling of active regions, and Section 4 presents the model comparison.

\section{Spectral libraries}

We note that the choice of libraries was based on public availability and therefore we by no means claim the following list to be complete. 

\subsection{\texttt{BT-Settl}}

 The BT-Settl models \citep{Allard2014} are based on the \texttt{PHOENIX}  code \citep{Hauschildt1996}. Spectra are calculated in 1D spherical geometry and are given only as a disc averaged flux. Since the models were developed also for ultra-cool dwarfs, the models include molecular lines (H2O, NH3, CH4, TiO, VO, CO, CO2, CN, FeH, CrH, MgH, CaH) and atomic linelists from combined databases (VALD, Kurucz Atomic Line Lists, NIST Atomic Spectra Database). Additionally, the model allows for 1D hydrodynamic cloud formation and settling which is required to accurately model ultra-cool dwarfs, although we note that the spectral types we consider in this work are too hot for clouds to form. 

\subsection{\texttt{PHOENIX NewEra}}

The \cite{Hauschildt2025} \texttt{NewEra} model grid  follows the setup of the commonly used \cite{Husser2013} 1D LTE stellar atmospheres library. The atmospheres are calculated with the \texttt{PHOENIX} code V20 in spherical geometry. The equation of state is described and solved through the Astrophysical Chemical Equilibrium Solver \citep[ACES;][]{Barman2011}, treating 839 different species (84 elements, 289 ions, 249 molecules, and 217 condensates). Atomic line opacities are taken from the Kurucz database \cite{Kurucz2005a, Kurucz2005b}. Molecular line opacities are primarily drawn from the ExoMol database \cite{Tennyson2016}, cut to $\sim$20.5 billion lines for the temperature range of the grid. Since the release of the Husser models over a decade ago, the molecular line lists have been substantially revised, in some scenarios introducing noticeable spectral differences, especially at optical wavelengths and for cool stars. These spectral differences arise despite minimal changes to the underlying atmospheric temperature–pressure structures. For the complete list of molecular lines, we refer to \citet{Hauschildt2025} Table 1 and Table 2. In the present work we utilised both the \texttt{PHOENIX-NewEraV2-LowRes-SPECTRA} which reach up to 1.6 microns, for extension to the Ariel wavelength regime we utilise the  \texttt{PHOENIX-NewEraV2-JWST-SPECTRA}. 

\subsection{\texttt{MPS-ATLAS}}

We utilise two different Merged Parallelised Simplified (MPS)- ATLAS \cite{Witkze2021}, spectral libraries in the context of the present work that use the same base radiative transfer (RT) code but rely on different constructions of the underlying atmosphere. The \texttt{MPS-ATLAS} code is  based on the ATLAS \cite{Kurucz1970} and DFSYNTHE \cite{Kurucz2005b, Castelli2005} codes, including an extension in its ability to treat optimum opacity distribution functions \citep{Cernetic2019}.

\cite{kostogryz2023mps} have employed the \texttt{MPS-ATLAS}  code to compute an extensive stellar atmosphere grid in plane-parallel geometry, and provided both specific intensity $I_\lambda(\mu)$ and disc-integrated values. Note that the disc-integrated values were calculated by disc-integrating the specific intensities.  The atomic and molecular line lists were taken from \cite{Kurucz2005b}.

More recently, \citetalias{}{Kostogryz2026} have presented a small grid of magnetized stellar atmospheres, where the atmospheric structures used in the RT calculations are computed with \texttt{MURaM}, which solves the conservative form of the MHD equations for a compressible, partially ionized plasma to simulate stellar atmospheres. In particular, it models convection self-consistently and is able to take into account initial magnetic field seeds which are important to simulate faculae.

\section{Forward model}

For the forward model we use two different approaches.
One is the classical TLSE effect, neglecting the spatial distribution of magnetic features and working with different temperatures for photosphere, faculae and spots.
The  spectrum $I(\lambda)$ can be written as

\begin{equation}
    I(\lambda) = I(\lambda)_{phot}\cdot\alpha_{phot} + I(\lambda)_{fac}\cdot\alpha_{fac} +I(\lambda)_{spot}\cdot\alpha_{spot},
\label{eq:tlse}
\end{equation} 

where $\alpha$ and $I(\lambda)$ are the disc filling factors and corresponding temperatures of the 3 distinct features, respectively. 
Another, though more complex approach is to model for a spotted (with spots and faculae) rotating photosphere built from the integration of the spectral contribution of a fine grid of surface elements. Such an approach is taken by the \texttt{StarSim} \citep{Herrero2016, Rosich2020} framework. While the current \texttt{StarSim}  version allows for combined modeling of photometric and spectral variability, we incorporate the extensions of the model by \cite{Porqueras2026} and Stucki in prep. which allow the calculation of the transmission spectrum and usage of the 3D models. A new version of the code, including those extensions will be released soon (Gomes et al., in prep.).

Similarly to Eq. \ref{eq:tlse}, one could write the resulting spectrum as  

\begin{equation}
    I^{jk}(\lambda) = I^{jk}(\lambda)_{phot}\cdot\alpha^{jk}_{phot} + I^{jk}(\lambda)_{fac}\cdot\alpha^{jk}_{fac} + I^{jk}(\lambda)_{spot}\cdot\alpha^{jk}_{spot},
\end{equation}

with $I^{jk}(\lambda)_{X}$ being the spectra of feature X either from the 3D library directly or the 1D \texttt{MPS-ATLAS} intensities with the corresponding temperatures from Tab. \ref{tab:stellar_parameters}  and $\alpha_{X}^{jk}$ being the disc coverage of feature $X$ at a surface element with coordinates $(j,k)$.

We allow for 1 to 10 magnetic features (1 features consists of spot and faculae pair) and for each number of features, we randomly construct 20 different surfaces, each with randomly drawn size distributions, longitudes, and latitudes. Spot sizes are drawn randomly, and a corresponding facula patch is assigned to each spot with an area equal to a randomly drawn integer multiplier between 1 and 3 times the spot area. A multiplier of 1 therefore corresponds to equal spot and facula coverage, while a multiplier of 3 means the facular area is 3 times larger than the associated spot. Since a multiplier of 1 results in a facular area equal to the spot area, the \texttt{StarSim} implementation currently defaults to discarding the faculae area. In that case, a new facular region is drawn independently with a size in the same range as the spot, ensuring that both components are always represented in the surface configuration.
This approach is used to calculate the \texttt{StarSim-3D} (meaning the combination of \texttt{StarSim} and the 3D-MHD spectra) and  \texttt{StarSim-1D} (meaning \texttt{StarSim} together with the 1D-\texttt{MPS-ATLAS}-CLV)  spectra. All calculations are based on a stellar inclination of 90$^\circ$.
The projected disc filling factors of spots $\alpha_{sp}$ and faculae $\alpha_{fc}$ obtained from each \texttt{StarSim} realisation are then inserted into Eq.~\ref{eq:tlse} together with the spectra from the \texttt{BT-Settl}, \texttt{MPS-ATLAS}, and  \texttt{NewEra} libraries at the corresponding component temperatures, yielding a disc-average contamination spectrum for each library and each surface configuration.

For the spectral types considered, we focus on the \texttt{3D MHD MPS-ATLAS} grid as our ‘ground-truth'. Consequently, we choose stellar atmospheres with similar effective temperatures, metallicities, and log g values for the other spectral libraries, assuming solar metallicity.
For the faculae component, we use the spectra computed for a surface magnetic field strength of 200G in  \citetalias{}{Kostogryz2026}, which is representative of the average field strength of solar faculae. The temperatures associated with these magnetized atmosphere calculations are listed in Table~\ref{tab:stellar_parameters}. The spot temperatures are calculated with respect to the photospheric temperature following \cite{Herbst2021}: 

\begin{equation}
    T_{spot} = -3.58 \cdot 10^{-5}T^2_{star} + 1.0188 \cdot T_{star} + 239.3
\label{eq:spot}
\end{equation}

where $T_{star}$ is taken to be the photospheric temperature. We note, that since there are no 3D MHD spot spectra available, we revert to using the 1D spectra in that case, noting that those spectra have known problems reproducing the contrast of the 3D spectra. There are no \texttt{MPS-ATLAS} library spectra below $T_\mathrm{eff} = 3500$K. For spot temperatures below this limit (M0 and M2 in Table~\ref{tab:stellar_parameters}), we rescale the 3500K spectrum by the ratio of the Planck function at the target temperature to that at 3500K, computed at each wavelength. While this preserves the overall spectral shape, it does not capture the changes in molecular band strengths at lower temperatures and should be regarded as an approximation. We emphasize that since Eq. \ref{eq:spot} is based on observations, the temperatures are a mix of the umbral and penumbral component that make up spots and we therefore do not distinguish between those two components in the present work.

\begin{table*}
\begin{center}
\begin{tabular}{c c c c c} 
\hline
Spectral type & Photospheric Temperature [K] & Spot Temperature [K] & Faculae Temperature [K] & log g \\
\hline
F3 &6800 & 4260 & 6880& 4.0\\
G2   &5780 &3950 & 5820& 4.4\\
K0   &5250& 3760&  5280& 4.4\\
K4   &4550& 3500 & 4570& 4.6\\
M0 &3750& 3120& 3750& 4.8\\
M2   &3560& 3030& 3520& 5.0\\
\hline
\end{tabular}
\end{center}
\caption{Spectral types and temperatures for the photosphere, faculae and spot component and log g considered in this work. Log g, photospheric and faculae temperatures were taken from \citetalias{Kostogryz2026} and correspond to their 'SSD' and '200G' case. The spot temperatures were calculated using Eq. \ref{eq:spot}}
\label{tab:stellar_parameters}
\end{table*}

\section{Results of the forward model exercise}

We generated roughly 200 spectra using the \texttt{StarSim} code for each spectral type considered. The heterogeneous stellar spectra are subsequently binned to the Ariel Tier 3 spectral resolution using the ArielRad radiometric simulator \citep{Mugnai2020} and \texttt{TauREx3} \citep{Al-Refaie2021, Al-Refaie2022}.

\begin{figure}
\centering
\includegraphics[width=\linewidth]
{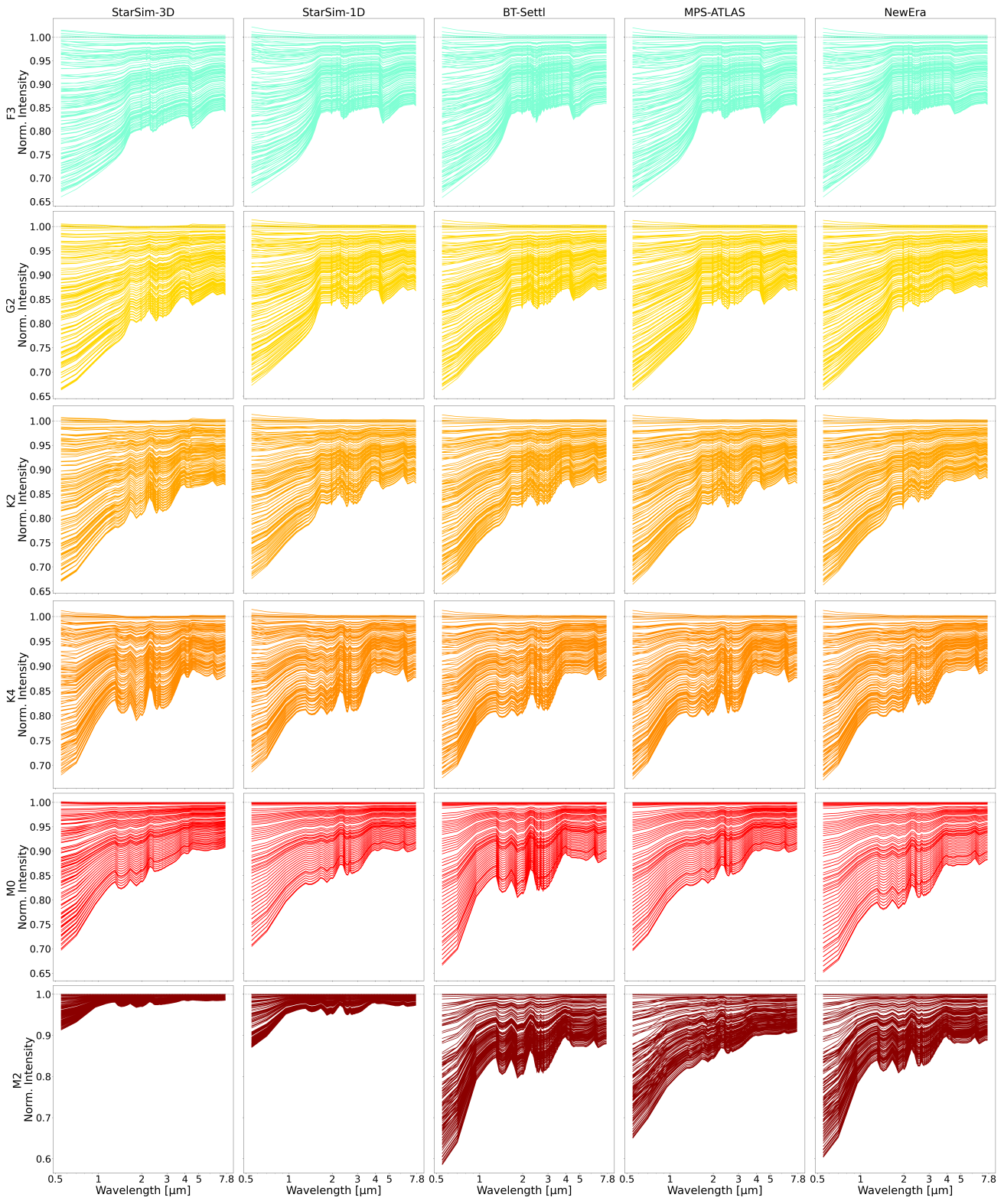}
\caption{Forward modelled stellar spectra as observed at Ariel Tier 3 spectral resolution with the different approaches. For more information on the different approaches representing each column we refer to the text.}
\label{fig:spectra}
\end{figure}

The set of spectra is shown in Fig. \ref{fig:spectra}. We normalized each spectrum to their respective photospheric spectrum. Intensities over 1 therefore correspond to an excess brightening of the star caused by faculae, dimming is observed when the spot spectral contribution exceeds that of the faculae.
Visually, some differences are clearly noticeable: for the G2 star, there is a steeper slope for higher intensities for all models compared to the 3D-MHD ones, while the calculations for the K-dwarfs appear relatively consistent. Once the cooler M-dwarfs are considered, differences become more pronounced. We quantify the differences later in this section.

\begin{figure}
\centering
\includegraphics[width=\linewidth]
{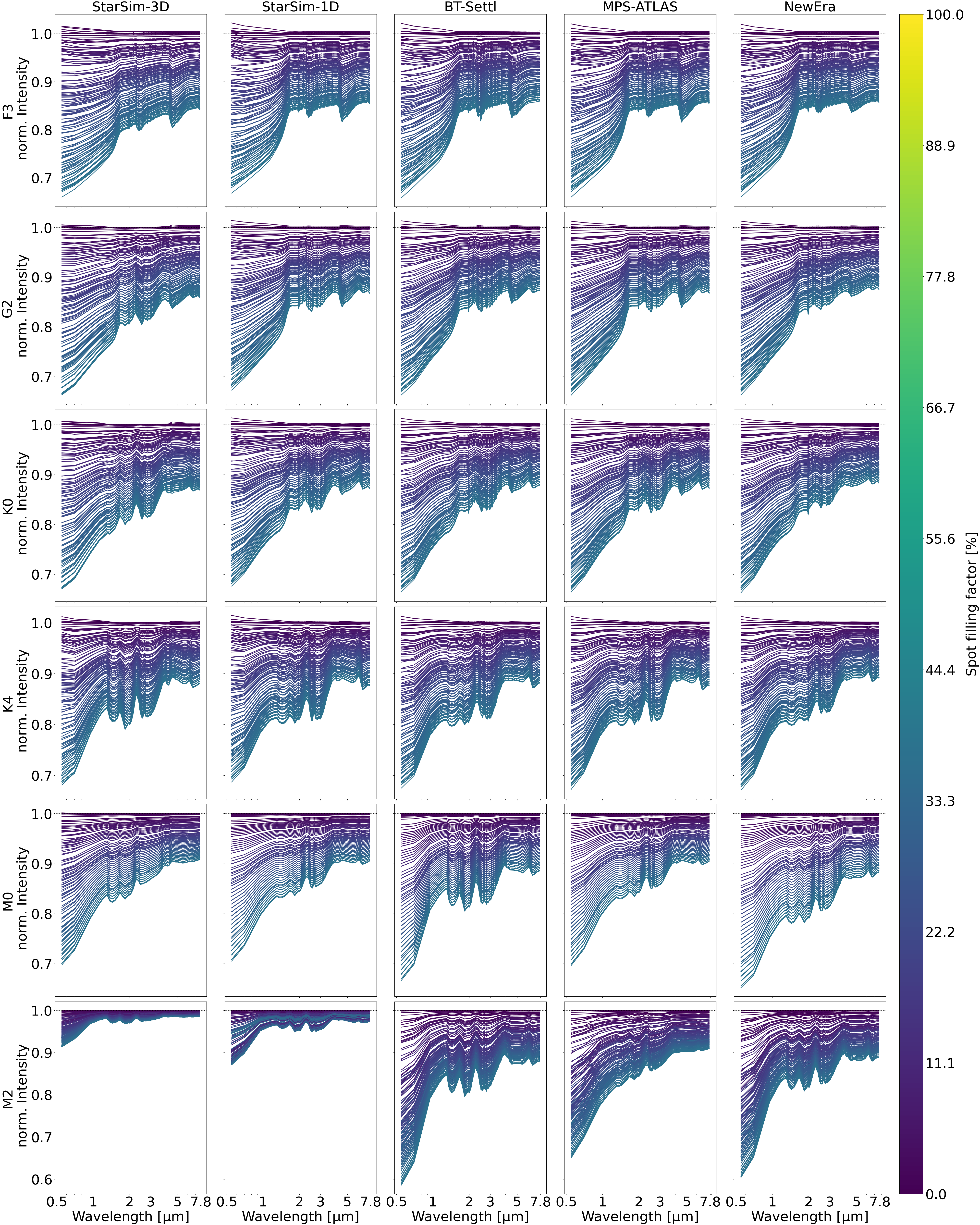}
\caption{Same spectra as shown in Fig. \ref{fig:spectra} with the projected spot area filling factors colorcoded. }
\label{fig:spectra_spots}
\end{figure}

\begin{figure}
\centering
\includegraphics[width=\linewidth]
{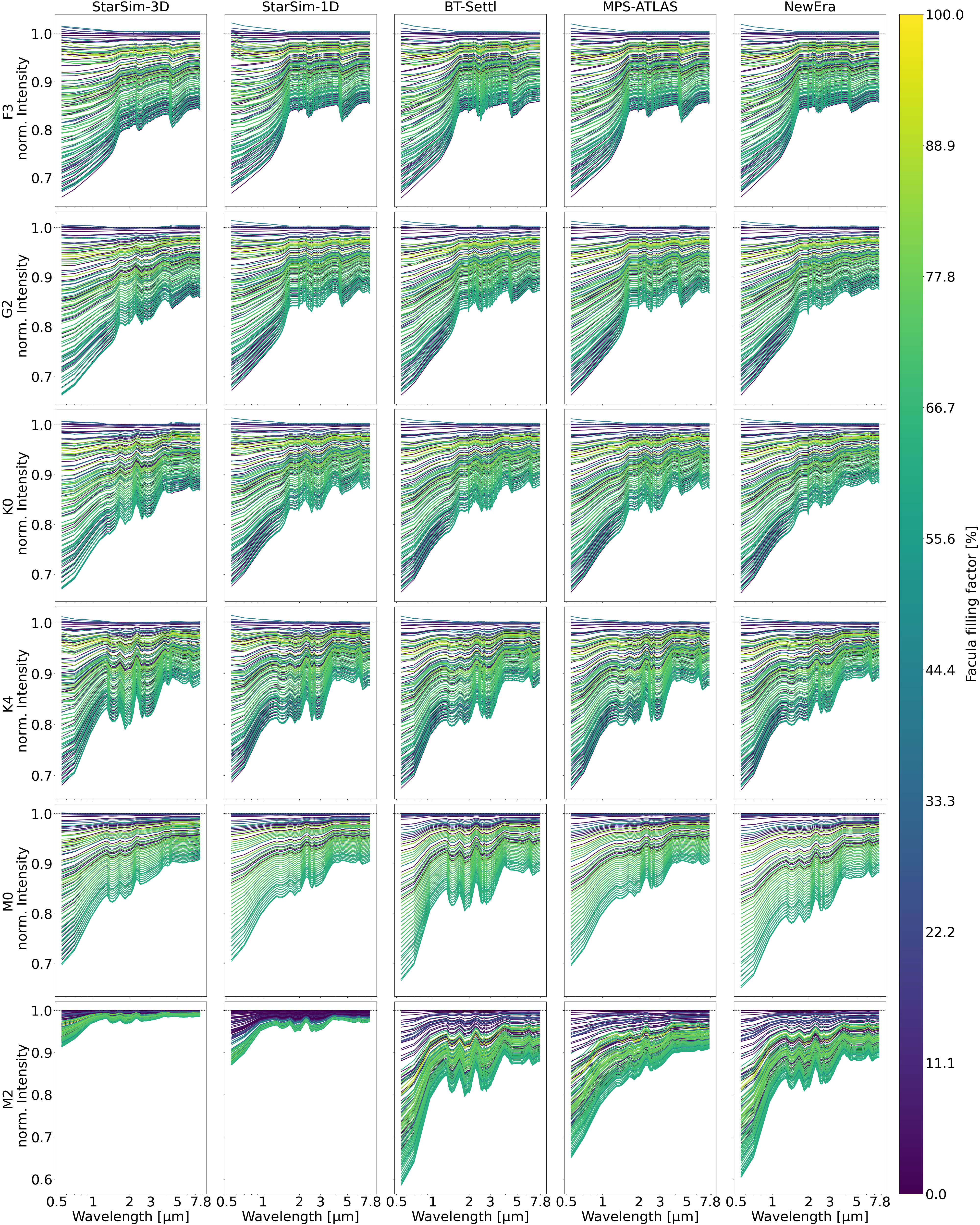}
\caption{Same as Fig. \ref{fig:spectra_spots} but for the projected faculae area filling factors.}
\label{fig:spectra_faculae}
\end{figure}

In Figs. \ref{fig:spectra_spots} and \ref{fig:spectra_faculae} and  we colourcode the disk coverage of spots and faculae respectively. The spot filling factors reach up to 40\%, and even up to 90\% for the faculae (mostly in cases where the spot filling factors are small). For all spectra, it is evident that with increasing spot coverage, the normalized intensity becomes smaller, meaning that there is net-darkening compared to the photosphere. However, if the faculae area coverages are considered, the trends are less clear. Generally, high faculae filling factors can counteract intermediate values of spot area coverages, and both a combination of low spot and low to intermediate faculae areas can lead to brightening with respect to the photosphere, specifically at wavelengths below 1 micron. In the case of the M0 and M2 calculations, high faculae filling factors also lead to net-darkening, which is counterintuitive, since ‘faculae' are usually associated with brightening. However, as discussed in \citetalias{Kostogryz2026} the effect of adding magnetic fields (which in this case are associated with faculae) becomes less pronounced for stars cooler than spectral classes of M0. While \citetalias{Kostogryz2026} presented this result in the form of CLV in dependence of $\mu$ for selected wavelengths, we calculate the contrast across the full Ariel wavelength regime. The contrast is defined as the difference between the photosphere and the considered magnetic region (faculae or spots), normalized to the photosphere. In this nomenclature, contrasts above 1 mean net-brightening with respect to the photosphere and below 1 darkening.

\begin{figure}
\centering
\includegraphics[width=\linewidth]
{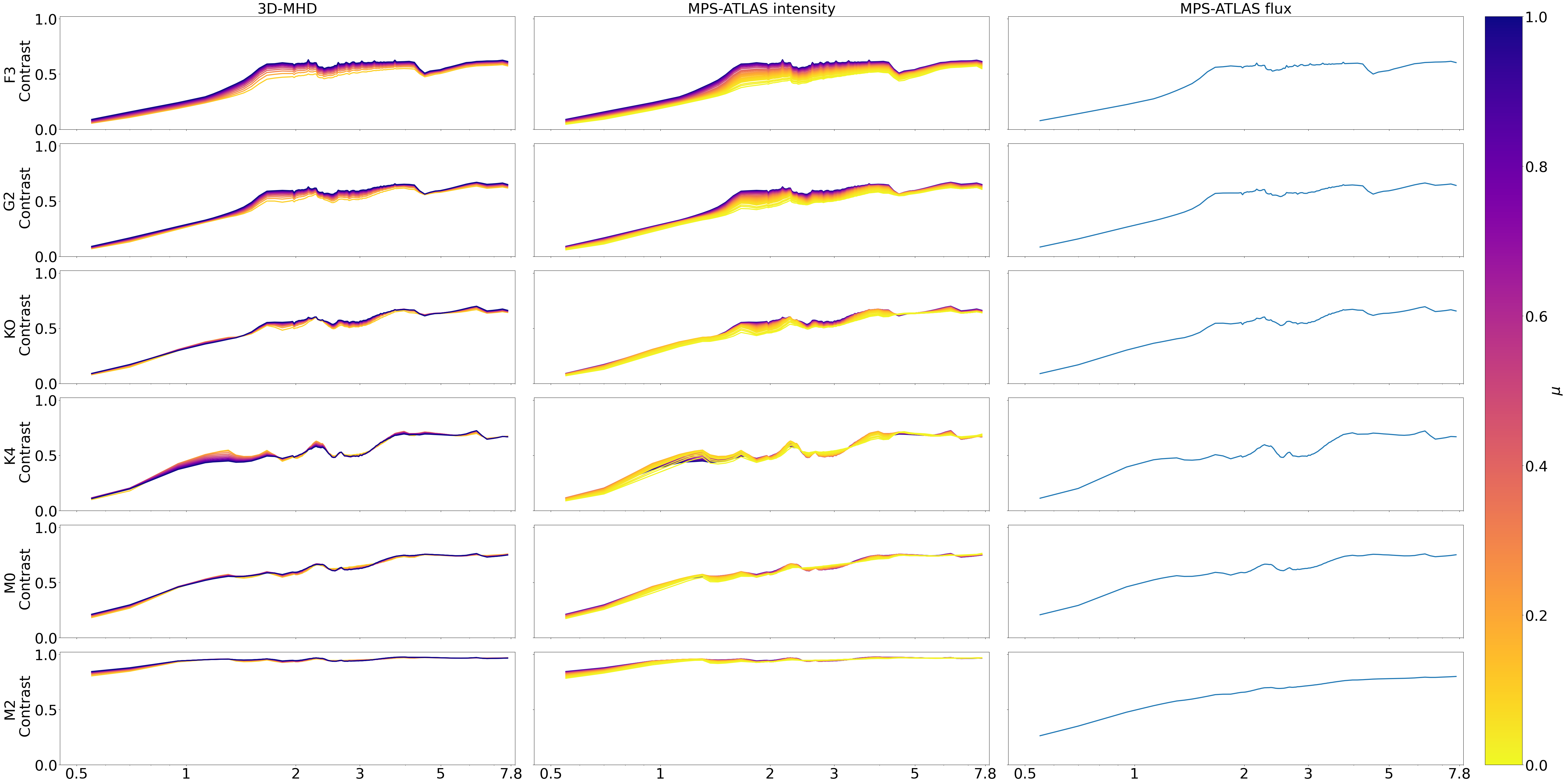}
\caption{Contrasts of the spot spectra to the photospheric spectra for 3 selected spectral libraries. Colorcoded is the $\mu$-angle for the \texttt{StarSim 3D MPS-ATLAS} spectra and the \texttt{StarSim 1D MPS-ATLAS CLV} spectra.}
\label{fig:contrasts_spots}
\end{figure}

\begin{figure}
\centering
\includegraphics[width=\linewidth]
{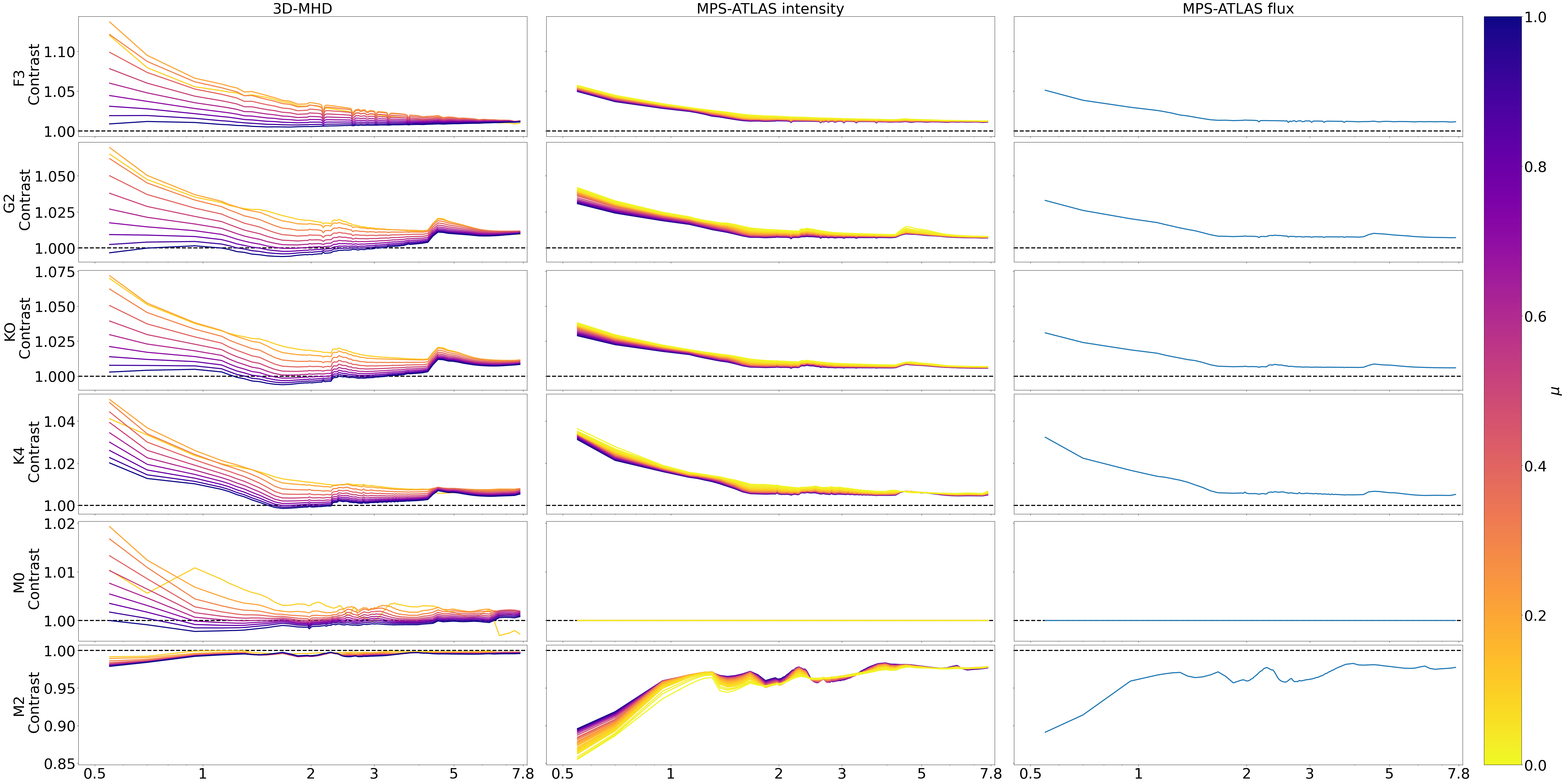}
\caption{Same as Fig. \ref{fig:contrasts_spots} but for the faculae contrasts.}
\label{fig:contrasts_fac}
\end{figure}

The contrasts are shown in Fig. \ref{fig:contrasts_spots} for the spots and Fig. \ref{fig:contrasts_fac} for the faculae. We note, that we only chose the \texttt{MPS-ATLAS} flux spectra in this comparison, and that the conclusions drawn from this experiment are the same for the \texttt{BT-Settl} and the \texttt{NewEra} spectra. We note, that the \texttt{MPS-ATLAS}-1D CLV spectra have a coarser $\mu$-grid compared to the \texttt{MPS-ATLAS}-3D spectra, particularly towards the limb, as the 3D spectra cut-off at $\mu$ = 0.1. The spot contrast is very similar, independent of whether a 3D or 1D quiet spectrum was used, and the \texttt{MPS-ATLAS}-flux spectra are most similar to the CLV spectra at $\mu$ = 0.5. This is consistent with the fact that the \texttt{MPS-ATLAS}-flux spectra were calculate from the \texttt{MPS-ATLAS}-1D CLV  spectra.
It is in the faculae component where the differences are striking. Again, the \texttt{MPS-ATLAS}-1D CLV flux spectra are similar to the $\mu$ = 0.5 values. Across the spectral types considered the \texttt{MPS-ATLAS}-3D faculae show a clear, non-linear trend with increasing contrast towards the limb, in particular with a strong slope shortward of 3 microns. As discussed in \citetalias{Kostogryz2026} the impact of 200G magnetic fields in the 3D calculations is decreasing with decreasing effective temperatures, with the smallest contrast still associated with brightening for the M0 calculations. Even though the resulting photospheric and facular temperatures are the same according to Tab. \ref{tab:stellar_parameters} shows that for M0,  there is still a difference in the contrast for the 3D calculations, which is because of the highly complex interplay of the magnetic fields in those simulations with both the granulation and the radiation field, which cannot be taken into account in 1D approaches. 

The brightening towards the limb is explained by the ‘hot wall' model \citep{Spruit1976}: magnetic flux concentrations cause a local depression of the ($\tau = 1$) surface, exposing hot lateral granulation walls. At disc centre the line-of-sight of the observer passes vertically through the depression, where the magnetic flux evacuates the flux tube (analogous to the process occurring within star spots), revealing cooler atmospheric layers. Towards the limb, the steeper viewing angle makes the hot walls increasingly visible, producing the characteristic brightening. Crucially, the brightening is a result of magnetic fields and geometry, both effects that cannot be captured by 1D models.  As discussed in \citetalias{Kostogryz2026} the magnetic suppression of convection dominates over the eﬀect of hot walls and faculae appear locally dark for stars below M0 spectral classes. The comparison presented in Fig. \ref{fig:contrasts_fac} highlights the need of including proper, physically motivated) faculae brightening (or in specific instances, dimming) in forward models of stellar contamination.

\begin{figure}
\centering
\includegraphics[width=\linewidth]
{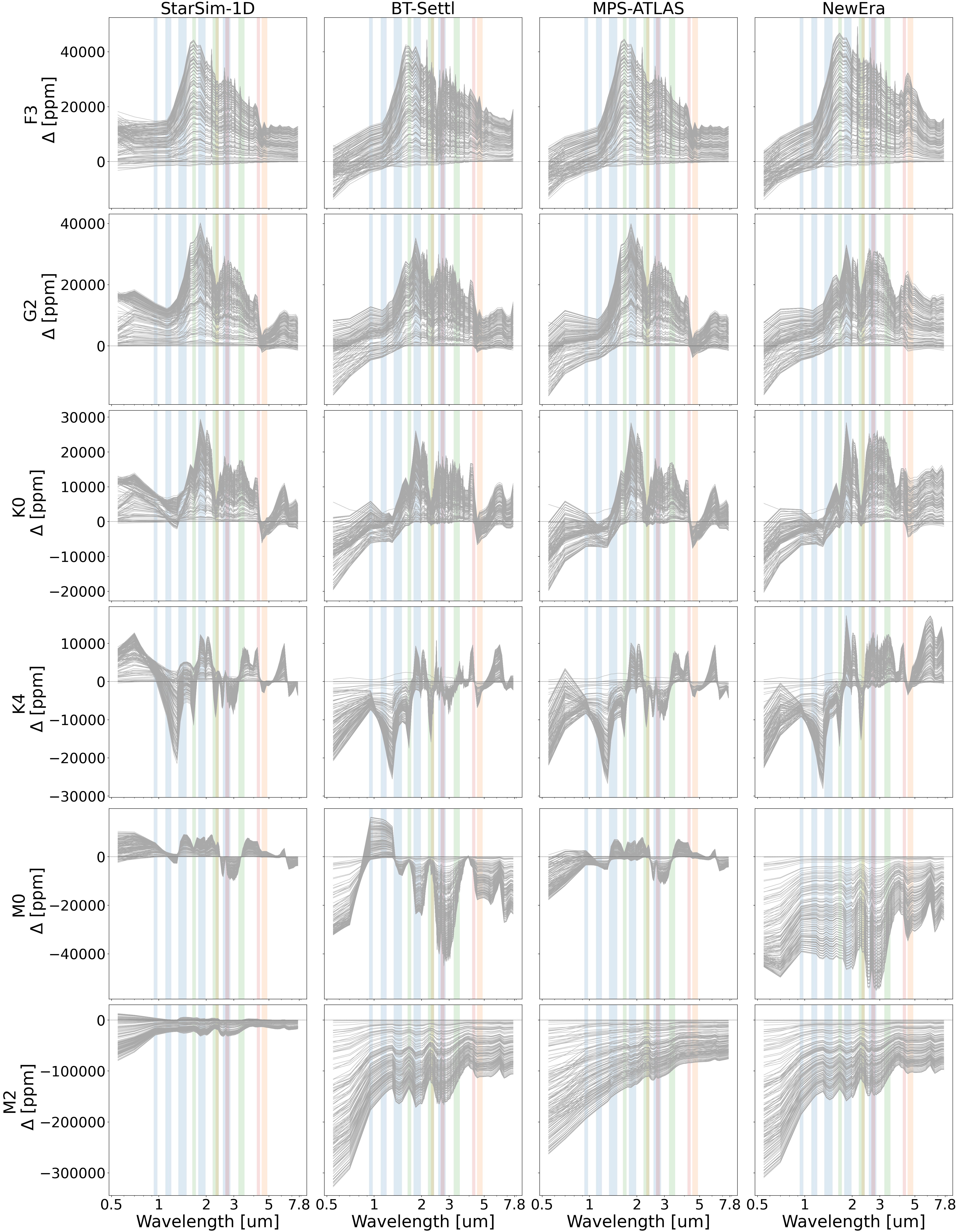}
\caption{Similar as Fig. \ref{fig:spectra} but this time we calculated the difference between the other libraries and the 3D-MHD spectra. The blue shaded vertical lines indicate H$_{2}$O, red shaded CO$_{2}$, orange shaded CO, and green shaded  CH$_{4}$ features.}
\label{fig:spectra_differences_features}
\end{figure}

Lastly, we emphasize the differences between the models by treating the  \texttt{MPS-ATLAS}-3D spectra as the ‘ground-truth' and calculate the differences between this and the other libraries. This is shown in Fig. \ref{fig:spectra_differences_features}, where we additionally added colored vertical lines for several molecular features ( blue: H$_{2}$O, red: CO$_{2}$, orange: CO, and green:  CH$_{4}$) that are predominately found in the atmospheres of both (sub) Neptunes and hot Jupiters, populations that represent the predominant targets of the Ariel mission.

\section{Conclusions and future work}

We have shown that across a range of stellar spectral types, from F3 to M2, it is insufficient to describe faculae as a spectrally hotter version of the photosphere. The facular contrast (defined as the ratio of the facular to photospheric spectrum) is not only highly wavelength dependent, but also shows a strong dependence on the $\mu$-angle, i.e. the position of the facula on the stellar disc, with both dependencies being functions of the stellar effective temperature.

Treating the \texttt{3D MHD MPS-ATLAS} spectra as the physical ground truth, the differences between those forward models spectra and those predicted by the 1D libraries (\texttt{BT-Settl}, \texttt{MPS-ATLAS-1D}, and \texttt{NewEra}) exceed the Ariel Tier 3 sensitivity threshold of $\sim$200 ppm across most of the wavelength range considered (0.5--7.8,$\mu$m), for most spectral types, and most activity levels considered. The differences are particularly pronounced for hotter stars and decrease towards cooler spectral types, with the M2 calculations showing the counter-intuitive result that faculae cause a net dimming rather than brightening. The differences between spectral libraries are large enough to mimic planetary atmospheric signals in several relevant bands (we considered H$_2$O, CO$_2$, CO, and CH$_4$). As noted, fixed limb-darkening coefficients introduce errors in the inferred planetary radius at the level of $\sim$3\%, and an inadequate limb-darkening law introduces wavelength-dependent biases exceeding 100 ppm \citep{Coulombe2024, Keers2024}. The library-dependent facular contrasts we find here are of comparable magnitude, and in several cases substantially larger, yet they arise from a modelling choice, namely the representation of faculae as a hotter photosphere, that is still routinely made without question. It is therefore crucial that both forward modelling exercises and machine learning approaches that rely on stellar models \citep[e.g.][]{Porqueras2026} to incorporate physically motivated facular CLV. The immediate next step, which forms the subject of Paper~II of this series, is to quantify the impact of model and spectral library mismatches on atmospheric retrievals, following a similar approach to \cite{Thompson2026}, and to assess the extent of any biases in retrieved atmospheric compositions, specifically those arising from the simplified treatment of faculae.

\section*{Acknowledgements}

This publication has been made possible by the spanish grants PID2021-125627OB-C31 funded by MCIU/AEI/10.13039/501100011033 and by “ERDF A way of making Europe”, PID2020-120375GB-I00 funded by MCIU/AEI, by the programme Unidad de Excelencia María de Maeztu CEX2020-001058-M, by the Generalitat de Catalunya/CERCA programme, by the SGR 01526/2021, the European Research Council (ERC) under the European Union’s Horizon Europe  programme (ERC Advanced Grant SPOTLESS; no. 101140786,  ERC Synergy Grant REVEaL grant No. 101118581) and the Marie Sk\l{}odowska-Curie Actions grant agreement No 101149286 (INCITE, and the ASI-INAFagreement2021-5-HH.2-2024.) A.T. acknowledges support from the ASI-INAF agreement 2021-5-HH.2-2024.

\section*{Conflict of Interest}

Authors declare no conflict of interest.

\section*{Data Availability}

This work has utilized publicly available open source software such as the \texttt{StarSim}(\url{https://github.com/dbarochlopez/starsim/tree/main}) and \texttt{Taurex3} (\url{https://github.com/ucl-exoplanets/taurex3}). The \texttt{PHOENIX NewEra} spectra are hosted by \url{https://www.fdr.uni-hamburg.de/record/17612}, the \texttt{MPS-ATLAS 1D} spectra by \url{https://edmond.mpg.de/dataset.xhtml?persistentId=doi:10.17617/3.NJ56TR} and the \texttt{MPS-ATLAS 3D} by \url{https://edmond.mpg.de/dataset.xhtml?persistentId=doi:10.17617/3.FBTIYY}.

\bibliographystyle{aasjournal}
\bibliography{main}

\end{document}